**Compositional tuning of ferromagnetism in $Ga_{1-x}Mn_xP$**

R. Farshchi[1,2], M. A. Scarpulla[1,2], P. R. Stone[1,2], K. M. Yu[2], I. D. Sharp[1,2], J.W. Beeman[2], H. H. Silvestri[1,2], L.A. Reichertz[2], E.E. Haller[1,2], and O.D. Dubon[1,2]

[1]Dept. of Materials Science and Engineering, Univ. of California, Berkeley, CA 94720
[2]Lawrence Berkeley National Laboratory, Berkeley, CA 94720

## Abstract

We report the magnetic and transport properties of $Ga_{1-x}Mn_xP$ synthesized via ion implantation followed by pulsed laser melting over a range of x, namely 0.018 to 0.042. Like $Ga_{1-x}Mn_xAs$, $Ga_{1-x}Mn_xP$ displays a monotonic increase of the ferromagnetic Curie temperature with x associated with the hole-mediated ferromagnetic phase while thermal annealing above 300 ºC leads to a quenching of ferromagnetism that is accompanied by a reduction of the substitutional fraction of Mn. However, contrary to observations in $Ga_{1-x}Mn_xAs$, $Ga_{1-x}Mn_xP$ is non-metallic over the entire composition range. At the lower temperatures over which the films are ferromagnetic, hole transport occurs via hopping conduction in a Mn-derived band; at higher temperatures it arises from holes in the valence band which are thermally excited across an energy gap that shrinks with x.

## Introduction

The ability to tune both electronic (charge) and magnetic (spin) functionalities in the same material is essential to the field of spin-based electronics, or spintronics [1]. Promising candidates for this purpose are III-Mn-V ferromagnetic semiconductors, where a few atomic percent Mn atoms occupy the group III (cation) sub-lattice [2]. The substitutional Mn atoms act as acceptors, and in the case of the archetypal ferromagnetic semiconductor $Ga_{1-x}Mn_xAs$, the holes they provide mediate the inter-Mn exchange. Despite extensive efforts to understand the nature of inter-Mn exchange, limited progress has been made in elucidating the interplay between hole transport and ferromagnetism in Ga-Mn-pnictide systems beyond $Ga_{1-x}Mn_xAs$. Here we report the dependence of magnetism and hole transport on composition in $Ga_{1-x}Mn_xP$ formed by ion implantation and pulsed-laser melting (II-PLM).

Ion implantation of a few atomic percent Mn into the III-V semiconductor lattice followed by pulsed-laser melting is a proven processing route for creating ferromagnetic semiconductors [3-6]. Rapid solidification of the molten implanted layer upon pulsed-laser irradiation traps the Mn atoms on substitutional lattice positions. The II-PLM process has been used to synthesize $Ga_{1-x}Mn_xAs$ films with magnetic and transport properties similar to films grown by low-temperature molecular beam epitaxy (LT-MBE) [3, 4, 6]. More recently II-PLM has been used to synthesize the carrier-mediated ferromagnetic phase in $Ga_{1-x}Mn_xP$ [5, 6]. An important difference between these two systems is the larger hole binding energy in $Ga_{1-x}Mn_xP$, namely 0.4 eV [7], compared to 0.11 eV in $Ga_{1-x}Mn_xAs$ [8]. One therefore expects holes to be more strongly localized in $Ga_{1-x}Mn_xP$, which may affect the ferromagnetic properties of this material.

**Sample Preparation**

GaP wafers (001) were implanted with $Mn^+$ ions over a dose range between $4x10^{15}$ $cm^{-2}$ and $2x10^{16}$ $cm^{-2}$ at an implantation energy of 50 keV. The wafers were then cleaved into approximately 5mm x 5mm square samples and irradiated with a single 0.4 $J/cm^2$ pulse from a KrF excimer laser ($\lambda$ = 248nm, FWHM = 18ns), homogenized to a spatial uniformity of ±5% by a crossed-cylindrical lens homogenizer. Upon laser irradiation, the amorphous implanted layer melts and regrows via liquid-phase epitaxy, hence restoring the crystallinity of the ion damaged layer. Finally, an extended HCl etch was carried out to remove a highly twinned surface layer leaving a single crystalline, single phase $Ga_{1-x}Mn_xP$ film [6].

Rutherford backscattering spectrometry (RBS) in combination with particle-induced X-ray emission (PIXE) analysis in the <110> and <111> axial channeling directions show that 70 to 85% of the Mn atoms occupy substitutional sites ($Mn_{Ga}$) and that the remaining Mn atoms do not occupy interstitial sites but instead are incommensurate with the lattice (i.e., at random locations). This level of substitutionality is not unlike $Ga_{1-x}Mn_xAs$ films of higher Mn concentration grown by LT-MBE, which contain on the order of 20% non-substitutional Mn [9].

The concentration profile of Mn in the resulting $Ga_{1-x}Mn_xP$ layer was measured by secondary ion mass spectrometry (SIMS). After processing, the $Ga_{1-x}Mn_xP$ films were characterized by a thickness in the range of 60-100 nm and a Mn concentration profile that reaches a maximum approximately 30 nm below the surface. We define x as the peak $Mn_{Ga}$ concentration. A maximum value of $Mn_{Ga}$ is given by the product of overall substitutional Mn fraction (from RBS/PIXE analysis) and maximum Mn concentration (from SIMS). We thus arrive at $Mn_{Ga}$ mole fractions (x) ranging from 0.018 to 0.042 for the samples studied.

## Results and Discussion

The dependence of ferromagnetic Curie temperature ($T_C$) on x as measured with a superconducting quantum interference device (SQUID) magnetometer is presented in Figure 1. Magnetization versus temperature was measured along a <110> in-plane direction in a field of 50 Oe (after saturation to 50 kOe), and data are shown for four compositions in Figure 1a. The Curie temperatures for these and other $Ga_{1-x}Mn_xP$ films made by II-PLM are plotted versus x in Figure 1b. Contrary to previous reports [10], $T_C$ increases monotonically with x. The linear relationship between $T_C$ and x observed for $Ga_{1-x}Mn_xP$ is not unlike that reported for $Ga_{1-x}Mn_xAs$ and described by mean-field models for hole-mediated exchange [11, 12]. This similarity between $Ga_{1-x}Mn_xP$ and $Ga_{1-x}Mn_xAs$ is consistent with the fact that the Mn $L_{3,2}$ X-ray magnetic circular dichroism in the two materials is nearly identical [13]. However, it is somewhat surprising given the more strongly localized nature of Mn in $Ga_{1-x}Mn_xP$ as discussed below. Linear extrapolation of $T_C$ to room temperature indicates that a composition x of approximately 0.18 may be required for room temperature ferromagnetism in this system.

Four-point sheet resistance measurements using pressed-on indium contacts in the van der Pauw geometry as a function of temperature are shown in Figure 2. The samples are insulating over the entire composition range. For samples with larger x (and higher $T_C$), there exists a distinct change in thermally activated transport between the low-temperature and the high-temperature resistivities. We attribute this to two distinct transport mechanisms [5]: i) one characterized by a smaller activation energy at lower temperatures corresponding to hole hopping in a Mn–derived band and ii) a larger activation energy mechanism associated with thermal excitation of holes from the Mn band to the valence band at higher temperatures. By comparison of the high temperature activation energies (slopes) as a function of composition, we find that the excitation energy for holes into the valence band decreases with increasing x. The charge hopping regime is not as apparent for the samples with lower x due to reduced inter-Mn wavefunction overlap leading to a strong increase in the activation barrier for hopping. In general the charge hopping temperature regime corresponds to that of the ferromagnetic state of $Ga_{1-x}Mn_xP$. Thus, holes mediating ferromagnetism in $Ga_{1-x}Mn_xP$ are strongly localized in contrast to the picture ascribed to $Ga_{1-x}Mn_xAs$ in which delocalized or weakly localized valence-band holes mediate ferromagnetism [11].

The presence of an excitation gap for localized holes in a Mn-derived band into the valence band is manifested in the infrared (IR) photoconductivity spectra of these films. In these measurements a constant bias is applied across the sample, and the electrical current through it is measured as a function of photon energy. As a result a spectrum of the photoconductive response of the sample is obtained. Figure 3 shows a photoconductivity spectrum measured at 5 K for $Ga_{1-x}Mn_xP$ with $x \sim 0.032$. The spectrum is characterized by a threshold in photoconductive response near 70 meV. Below this edge there is no signal as has been confirmed by the measurement of the far IR photoconductivity spectrum using a Fourier transform spectrometer with greater sensitivity in this photon energy range (inset of Figure 3). Similar experiments were carried out previously for $x \sim 0.042$ and yielded a photoconductivity response threshold at a considerably lower energy of 26 meV [5]. The compositional dependence of both the threshold and the activation barrier for hole transport paints a picture of holes localized in a Mn-derived band that is separated from the valence band by an energy gap that shrinks with increasing x.

Similar to $Ga_{1-x}Mn_xAs$, carrier-mediated ferromagnetism in $Ga_{1-x}Mn_xP$ is not stable against thermal annealing above 300 ºC [9, 14]. Figure 4a shows magnetization versus temperature for samples annealed at 300 ºC - 400 ºC for 15 minutes in flowing $N_2$ gas. These measurements show a significant decrease of $T_C$ upon thermal annealing, which as reflected by PIXE measurements is accompanied by a relocation of a fraction of Mn atoms from substitutional to incommensurate (random) sites.

Figure 4b presents $T_C$ versus x. The tuning of x is achieved by variation of the implantation dose (as presented in Figure 1) and by thermal annealing of samples having an initial (prior to annealing) $Mn_{Ga}$ mole fraction of x ~ 0.034. The stronger dependence of $T_C$ on x for the series of annealed samples suggests that relocation of $Mn_{Ga}$ to incommensurate sites leads to the partial compensation of the remaining $Mn_{Ga}$ or to the formation of defect complexes that render a fraction of $Mn_{Ga}$ ferromagnetically inactive. We note that a similar effect has been observed in $Ga_{1-x}Mn_xAs$ films grown by LT-MBE [9]. Finally, the thermal instability of $Mn_{Ga}$ in $Ga_{1-x}Mn_xP$ at temperatures above 300 ºC demonstrates that any high-temperature ferromagnetism in Mn-containing GaP that has been subjected to extended exposure to such temperatures is unlikely to originate from the dilute $Ga_{1-x}Mn_xP$ (alloy) phase.

**Summary**

We have determined the compositional dependence of $T_C$ in $Ga_{1-x}Mn_xP$. The films are non-metallic for x ranging from 0.018 to 0.042. and hopping conduction is observed at temperatures over which the films are ferromagnetic. At higher temperatures the sheet resistance is governed by holes that are thermally excited into the valence band across an energy gap that shrinks with increasing x. Finally, $Mn_{Ga}$ in $Ga_{1-x}Mn_xP$ is not stable to thermal annealing above 300 °C, which causes a strong decrease of $T_C$ accompanied by a relocation of a fraction of $Mn_{Ga}$ to incommensurate sites.

**Acknowledgements**

This work is supported by the Director, Office of Science, Office of Basic Energy Sciences, Division of Materials Sciences and Engineering, of the U.S. Department of Energy under Contract No. DE-AC02-05CH11231. P.R.S. acknowledges support from an NDSEG Fellowship.

**Figure Captions:**

**Figure 1**. a) Magnetization versus temperature for x ~ 0.018 (circles), 0.029 (squares), 0.034 (diamonds), and 0.038 (triangles). Hysteresis loops for x ~ 0.034 measured at 5K (filled squares) and 30K (open squares) are also shown (inset).

b) Variation of $T_C$ with x. Measurements for additional samples are included.

**Figure 2**. Dependence of sheet resistance on temperature for x ranging from 0.018 to 0.042. Distinct regimes of thermally activated transport are observed for samples with higher values of x (and larger $T_C$).

**Figure 3**. Main panel: IR photoconductivity spectrum from a film with x ~ 0.032. The photoconductivity threshold near 70meV is confirmed by the photoconductivity spectrum measured using a far IR Fourier transform spectrometer (inset). Features in the spectrum associated with the instrument response have not been removed.

**Figure 4**. a) Temperature dependence of bulk magnetization for II-PLM samples with x ~ 0.034 after (post-PLM) thermal annealing for 15 minutes at various temperatures. The sample annealed at 400 ºC shows near complete loss of ferromagnetism. b) Dependence of $T_C$ on x. The $Mn_{Ga}$ mole fraction is tuned by varying the implantation dose (squares) and the annealing temperature (triangles).

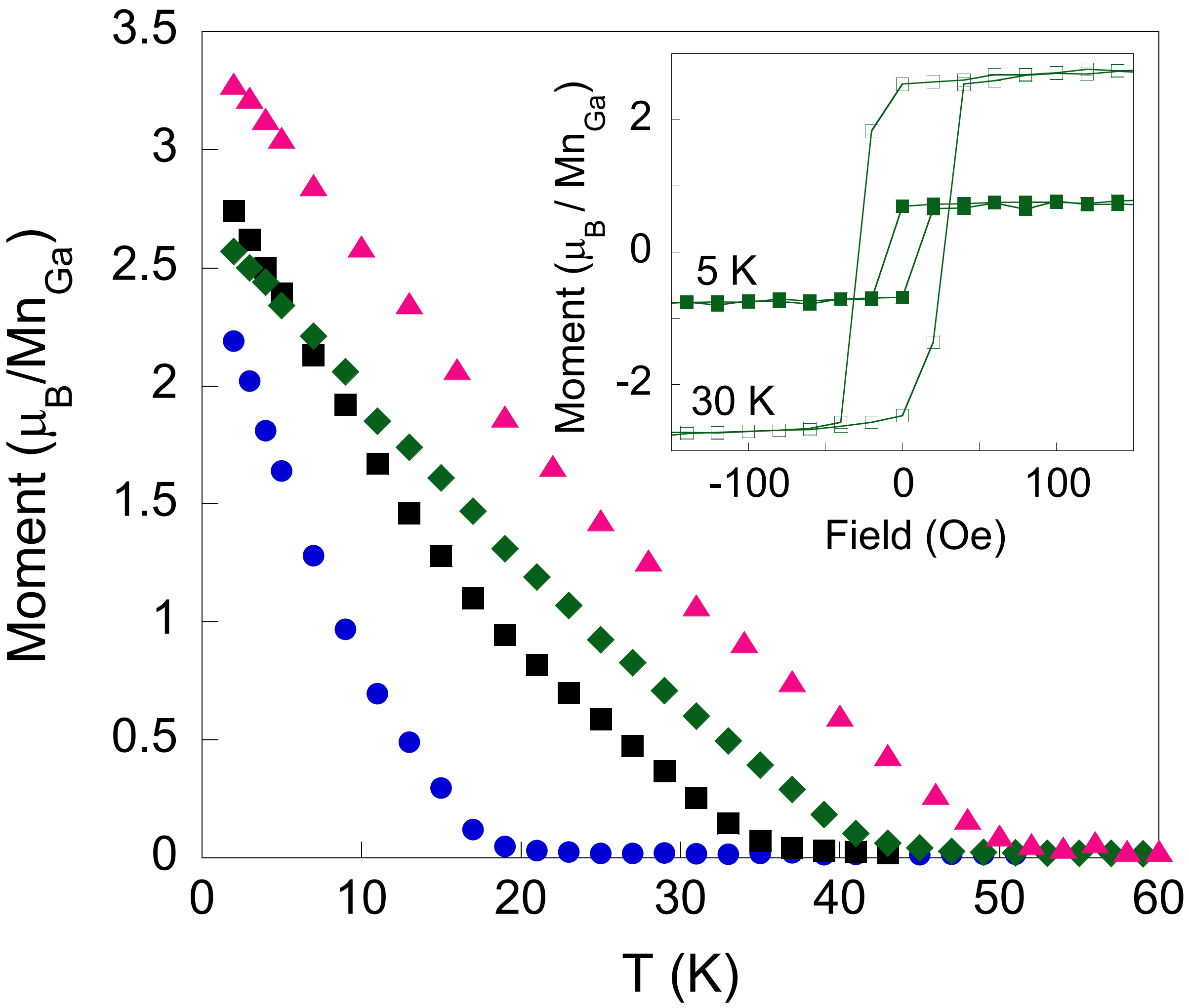

Figure 1a

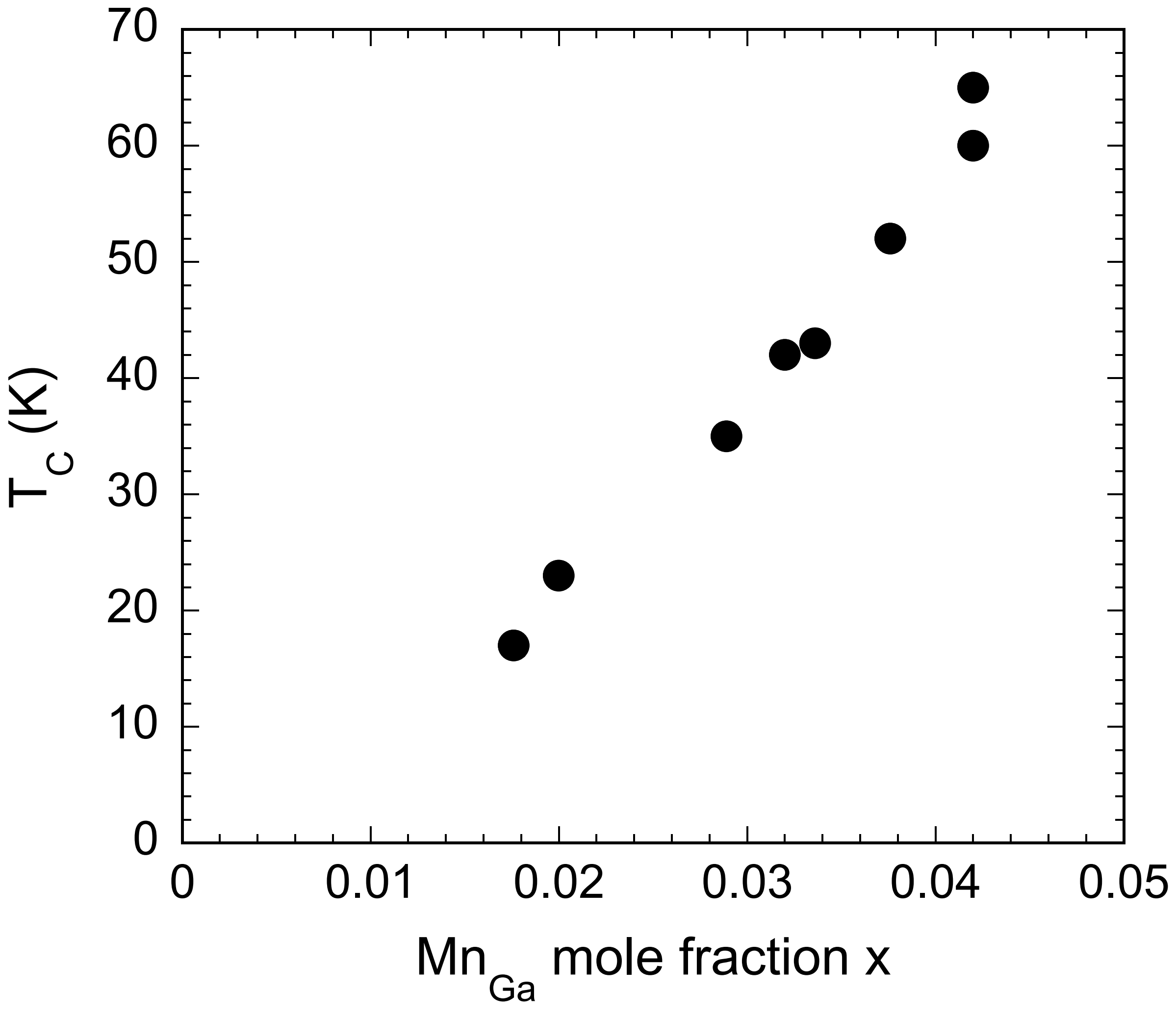

Figure 1b

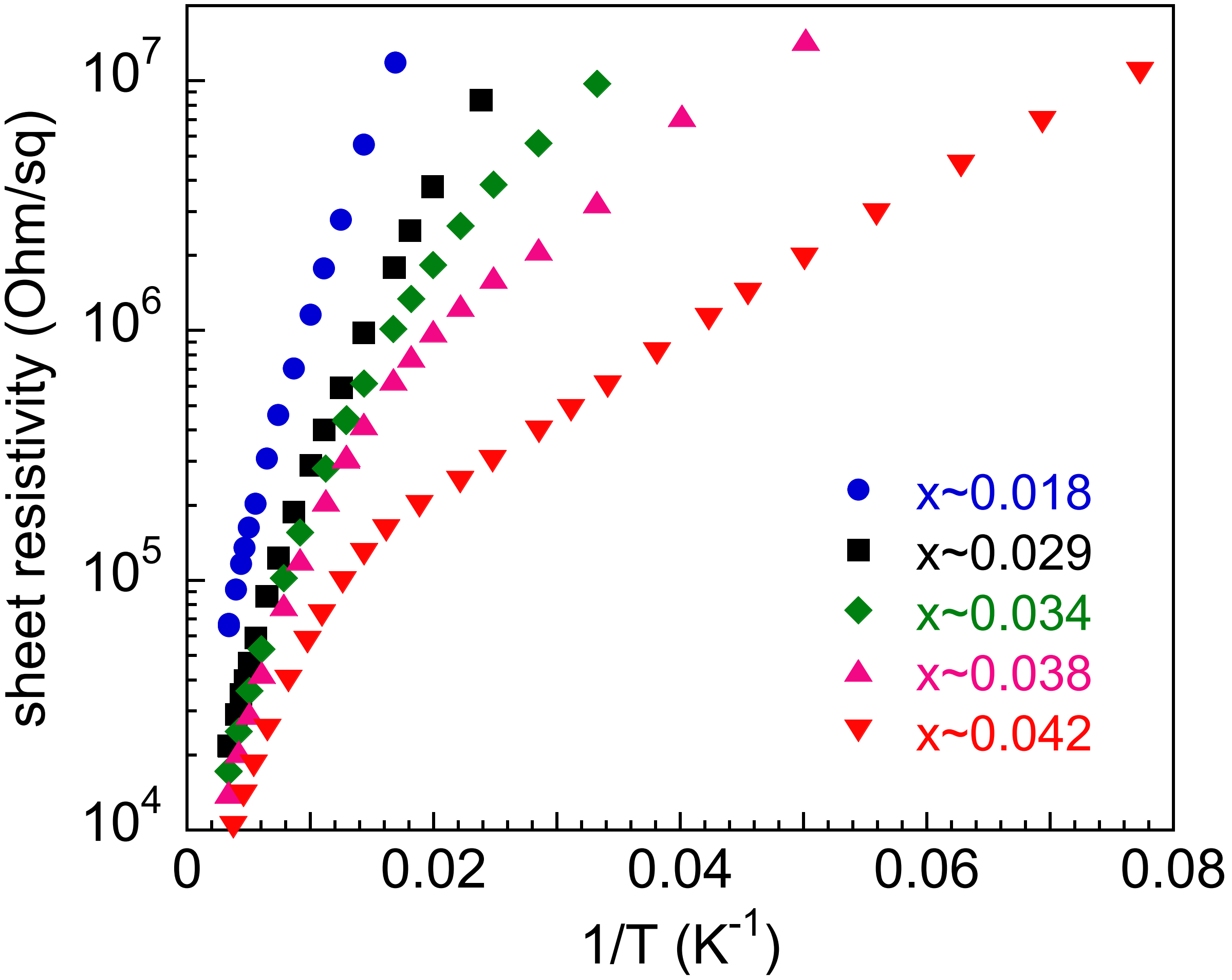

Figure 2

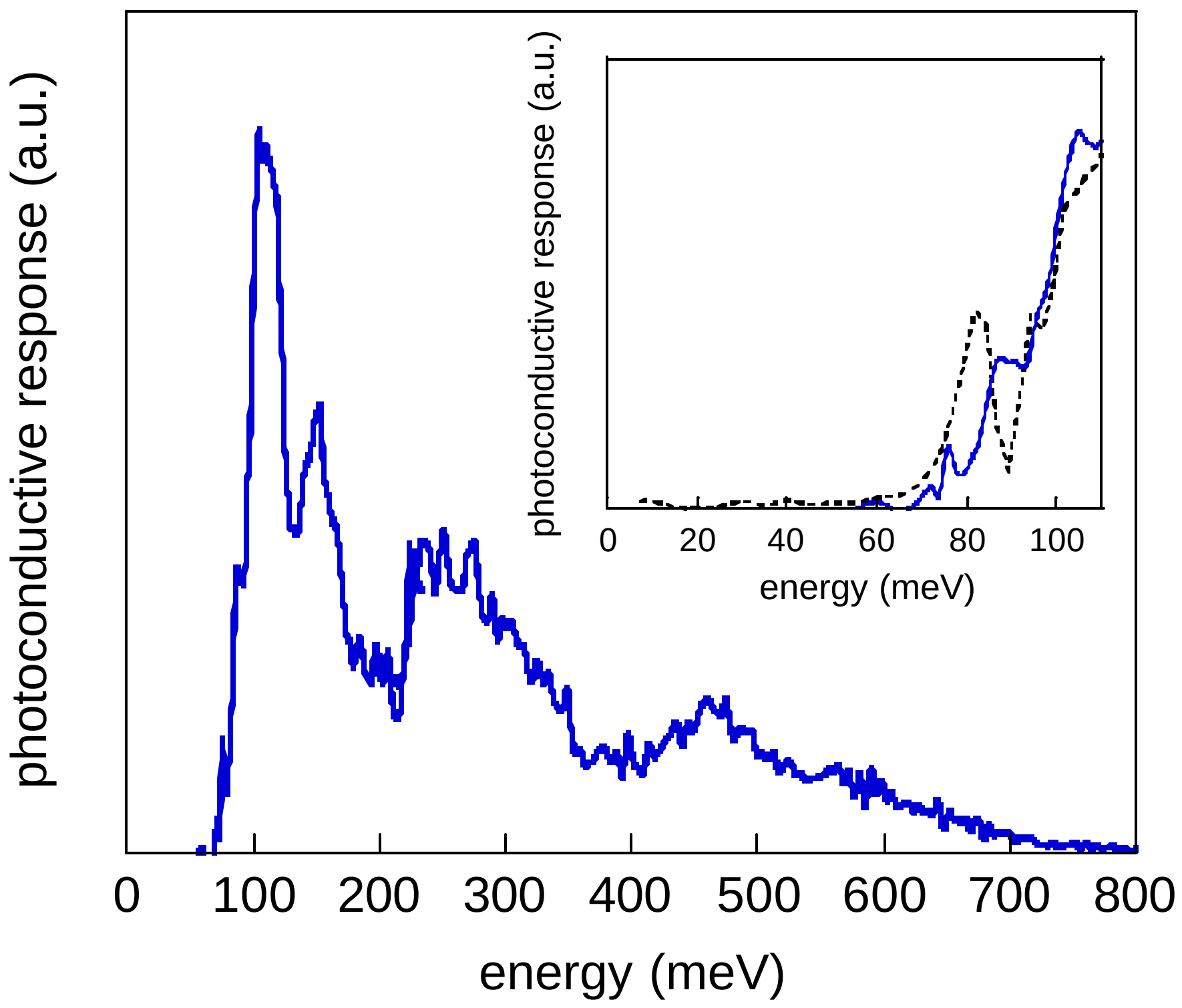

Figure 3

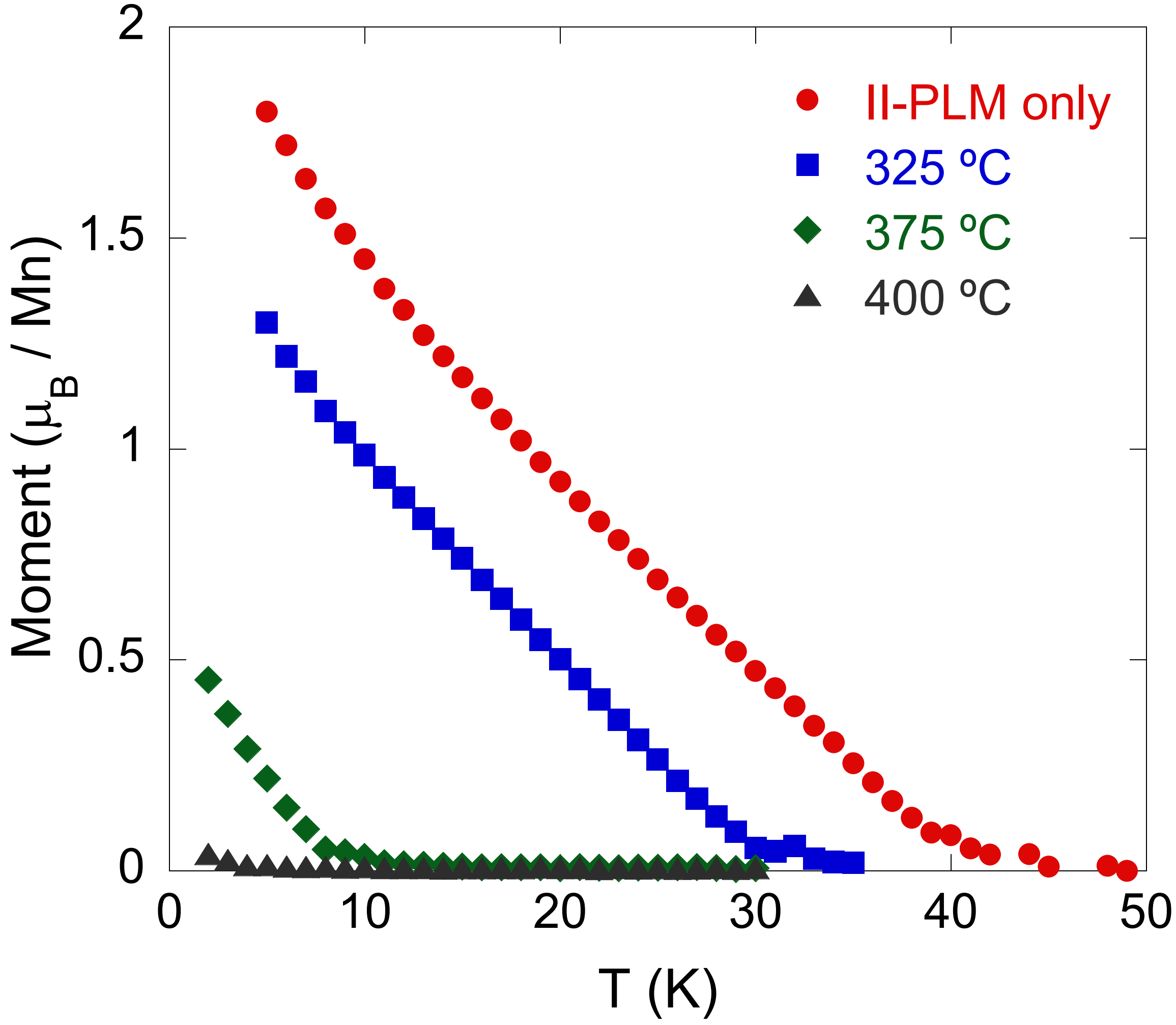

Figure 4a

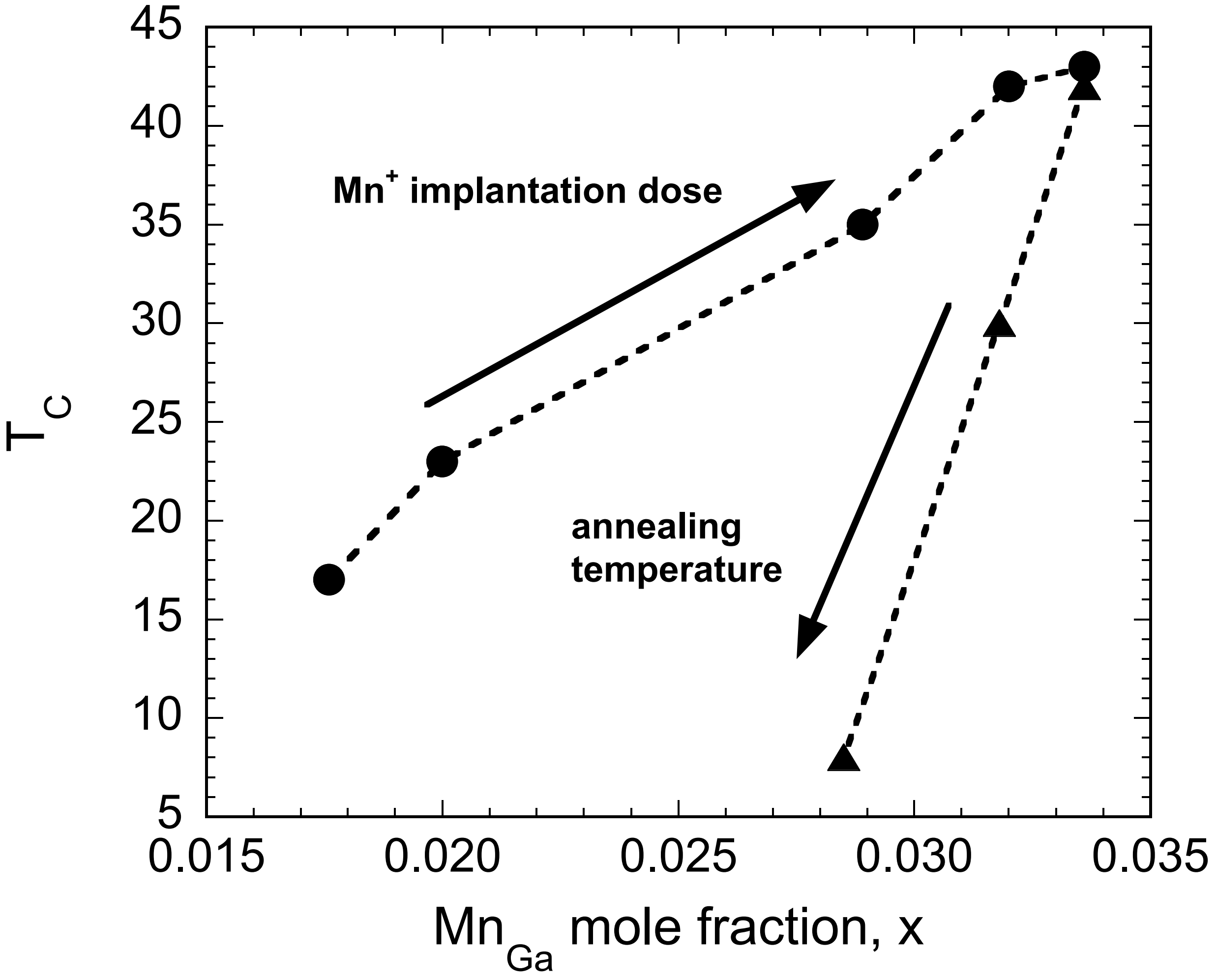

Figure 4b